\documentclass{article}
\usepackage{graphicx}
\usepackage{textcomp}
\usepackage{xcolor}
\usepackage{comment}
\usepackage{authblk}
\usepackage{url}

\date{}

\title{Medication counseling with large language models: balancing flexibility and rigidity}

\author{
Joar Sabel\textsuperscript{1}, 
Mattias Wingren\textsuperscript{2  \thanks{Corresponding mail: mattias.wingren@abo.fi}} ,
Andreas Lundell\textsuperscript{3},\authorcr
Sören Andersson\textsuperscript{2},
Sara Rosenberg \textsuperscript{4},
Susanne Hägglund\textsuperscript{2}, \authorcr
Linda Estman \textsuperscript{5},
Malin Andtfolk\textsuperscript{4} \\

\textsuperscript{1}  Department of Engineering and Information Technology\\
Åbo Akademi University, Vaasa/Turku, Finland\\
\textsuperscript{2}Experience Lab, Åbo Akademi University, Vaasa, Finland\\
\textsuperscript{3}Department of Engineering and Information Technology\\
Åbo Akademi University, Vaasa, Finland\\
\textsuperscript{4}Department of Natural and Health Sciences\\
Åbo Akademi University, Vaasa, Finland\\
\textsuperscript{5}Department of Caring and Ethics\\
University of Stavanger, Stavanger, Norway\\
}

\begin{document}

\maketitle

\begin{abstract}
The introduction of large language models (LLMs) has greatly enhanced the capabilities of software agents. Instead of relying on rule-based interactions, agents can now interact in flexible ways akin to humans. However, this flexibility quickly becomes a problem in fields where errors can be disastrous, such as in a pharmacy context, but the opposite also holds true; a system that is too inflexible will also lead to errors, as it can become too rigid to handle situations that are not accounted for. Work using LLMs in a pharmacy context have adopted a wide scope, accounting for many different medications in brief interactions --- our strategy is the opposite: focus on a more narrow and long task. This not only enables a greater understanding of the task at hand, but also provides insight into what challenges are present in an interaction of longer nature. The main challenge, however, remains the same for a narrow and wide system: it needs to strike a balance between adherence to conversational requirements and flexibility. In an effort to strike such a balance, we present a prototype system meant to provide medication counseling while juggling these two extremes. We also cover our design in constructing such a system, with a focus on methods aiming to fulfill conversation requirements, reduce hallucinations and promote high-quality responses. The methods used have the potential to increase the determinism of the system, while simultaneously not removing the dynamic conversational abilities granted by the usage of LLMs. However, a great deal of work remains ahead, and the development of this kind of system needs to involve continuous testing and a human-in-the-loop. It should also be evaluated outside of commonly used benchmarks for LLMs, as these do not adequately capture the complexities of this kind of conversational system.
\end{abstract}


\maketitle

\section{Introduction}

Software agents have historically been used in a wide range of contexts, from cybersecurity \cite{alluhaybi_survey_2019} to medicine \cite{isern_agents_2010}. Some of these agents have been designed to interact directly with humans, such as the well-known chatbot \cite{calvaresi_exploring_2023}. Traditionally, these systems have been rooted in a rule-based architecture, making it difficult to achieve natural conversations. Often, the human partner would need to adapt to the system and not vice versa \cite{pelikan_why_2016}.

With the introduction of large language models (LLMs), agents have been given tools to better handle the complexities of human conversation. Since LLMs are probabilistic, they are no longer constrained to predefined rules. However, this flexibility causes issues on the other end of the spectrum, as LLM systems have a tendency to hallucinate --- that is, generate incorrect or unrealistic responses in relation to their source content \cite{ji_survey_2023}. A factor that affects a model's tendency to hallucinate is its context window. The context window of an LLM is the amount of text the model can process at once and can effectively be thought of as its working memory. However, this memory comes with serious limitations. When contextual information increases, relevant facts are more likely to be ignored, especially information in the middle of the context window \cite{liu_lost_2023}, similar to humans \cite{gobet_chunking_2001}. What is unique to LLMs, however, is how hallucinations work in tandem with limitations in their contextual understanding. The greater the contextual information, the more likely the model is to hallucinate, for example by being overly confident or simply disregarding relevant facts of a given situation \cite{huang_survey_2025}.

These flaws can have consequences of varying degrees depending on the context, and one context in which such flaws can be costly, both in terms of health and financially, is medication counseling \cite{kohn_errors_2000, van201117}. Despite these flaws, the flexibility and generative capabilities of LLMs are of interest to the field. Liu et al. \cite{liu_pharmacygpt_2024}, for example, propose a framework for emulating the role of clinical pharmacists, whereas Osheba et al. \cite{osheba_leveraging_2025} use LLMs to verifying prescriptions and answer inquiries. These, and other work, \cite{shin_performance_2024, van_rx_2024} handle many different medications during a short interaction; but what also needs to be studied is medication counseling concerning one specific medication during a longer time. The result will vary between these two approaches since a longer task with more conversational requirements gives rise to opportunities for different kind of errors; but it is also important to examine since this kind of interaction is more akin to medication counseling done by humans. Our project starts from this other end, building a system that aims to provide satisfactory medication counseling in a specific case: counseling regarding emergency contraceptive pills (ECPs).

\section{Background}

This work is grounded in a project aiming to contribute interdisciplinary knowledge on how social robots can support and promote medication safety by providing medication counseling at pharmacies \cite{andtfolk_further_2022}. What is of interest is a robot's ability to provide counseling regarding ECPs. As mentioned before, we see it as a fruitful strategy to begin studying counseling regarding one specific case, and this one seems appropriate since it is a medication that does not need a prescription, but still requires additional counseling according to Finnish law \cite{fimea_lisaneuvontaa}. Since ECPs can be stressful to purchase \cite{gonzalez-mesa_anxiety_2019}, we also see the interaction with an agent as a potentially less unconformable interaction than one with a human pharmacist.

The procedure of providing counseling regarding ECPs follows a rigid procedure \cite{fimea_lisaneuvontaa}. The project has studied this process in detail and divided the task into goals and sub-goals \cite{wingren_using_2025}. The goals and sub-goals together number 35, but the following five goals illustrate the procedure quite well: (1) check time since intercourse, (2) check for contraindications, (3) present side effects and the underlying mechanism of the pills, (4) offer choices, and (5) offer advice. Among these, (2) is especially crucial since it ensures the safety and effectiveness of the ECPs; but this goal is also difficult, because the system needs to be able to interpret potential contraindications (i.e., allergies, diseases and medication) that can negatively interact with specific ECPs. This would be trivial if customers described contraindications like these in a uniform way, but naturally they might use varying terms for the same condition, making interpretation problematic. 

At the beginning of the project, the system meant to provide medication counseling was based on a state graph, and therefore its conversational capabilities were rigid and limited. Such a system is by design constrained to such a degree that it quickly loses its usefulness. Introducing LLMs into the system provides flexibility that is needed to handle these difficult situations; but as mentioned before, this flexibility needs to be tempered.

\subsection{Tools} 
\label{tool-agents}

A way to introduce rigidity to an LLM system is by using tools \cite{yuan_easytool_2024}, external functions such as application programming interface (API) calls or code execution which can be used to improve the quality and flexibility of their responses. For example, in our case all contraindications can be accessed through a tool call which avoids polluting the LLMs context when the information is not needed; in essence it is a form of retrieval augmented generation (RAG).

Tools become even more useful when LLMs are used to reason about which tool to call and subsequently interpret the output of said tool and act based on it; this type of agent is called a ReAct agent.

\subsection{ReAct Agents}
\label{react}
 
ReAct agents (Reason + Act) were introduced by Yao et al. \cite{yao_react_2023}, and unlike traditional task-solving pipelines that separate reasoning and execution, these agents can interact with external environments (e.g., APIs) while maintaining a structured thought process. In contrast to, for example, chain-of-thought agents \cite{wei_chain--thought_2023} which only reason, ReAct agents combine reasoning with action, improving their decision-making and reducing hallucination.

However, the more responsibilities an agent is given, the higher the risk of it performing unwanted behavior. One way of mitigating this is by dividing the task between multiple agents.

\subsection{Multi-agent Systems}
In contrast to a single-agent system (SAS) which often struggles with longer inputs and more complex tasks due to its monolithic nature \cite{li_long-context_2024}, a multi-agent system (MAS) can distribute both context and reasoning between several agents. Task decomposition \cite{khot_decomposed_2023} can be leveraged to split the main task between different agents as subtasks, thus reducing the ``cognitive'' load of each agent. Task decomposition can be thought of in the same way as humans break down tasks into smaller steps. Decomposing tasks leads to better response quality since each agent operates within a narrower problem space. Another advantage of an MAS is its modular composition, which makes it easier to change the system without impacting its overall functionality. Finally, utilizing shared memory between all agents enables them to cross-check each other's output, which allows the agents to attempt self-correction whenever an error occurs, since they are able to reason about what has happened and what should happen next. Self-correction, in this case, refers to the agent being able to determine that something has gone wrong and to attempt to take action in order to remedy the situation.

\section{Implementation}
The system is constructed as an MAS consisting of three agents, the (1) Conversationalist, (2) Medicine interpreter and (3) Symptom assessor, all being ReAct agents. The system uses a graph-based implementation, where the agents are represented as nodes and their interaction paths are represented as edges.

\begin{figure*}
    \centering
   \includegraphics[width=\linewidth, height=20cm, keepaspectratio]{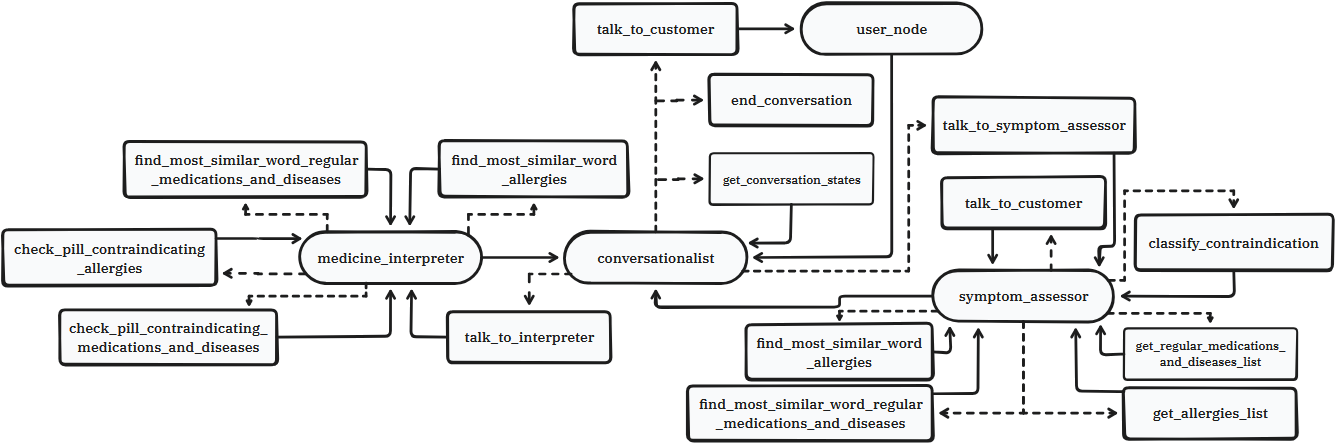}
    \caption{Graph representation of the system. Optional paths are dashed lines. Mandatory paths are solid lines. Agents are ellipse-shaped, and tools are box-shaped.}
    \label{fig:sys-grpah}
\end{figure*}

\begin{figure}
  \centering
  \includegraphics[width=1\linewidth]{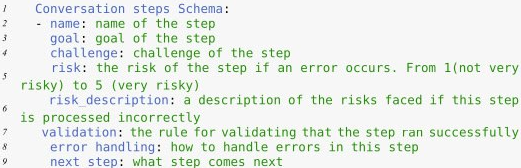}
  \caption{Structure of the conversation specification steps}
  \label{fig:yaml}
\end{figure}

As mentioned before, the task of providing counseling regarding ECP was studied in great detail. The analysis used was a hierarchical task analysis \cite{stanton_hierarchical_2006}, and since it decomposes a task into smaller parts, it naturally fits our system that uses task decomposition. We have also developed an extension of this method for more fine-grained analysis. Part of this analysis was used for the specification regarding the required content of the conversation. The specification details what must be discussed with the customer during the conversation and what information must be obtained and considered in the process of determining viable medication. The specification was converted into the YAML format, as it can easily be interpreted by LLMs due to its rigid token structure and hierarchical organization. The structure of the YAML can be seen in Figure \ref{fig:yaml}. The project also provides a knowledge base detailing the composition and contraindications of the ECPs. The specification and the knowledge bases provides criteria to determine if an interaction is a success or a failure; if all the required topics have been discussed and the customer has received the right set of ECPs, it has been successful, otherwise not. Ideally, both the knowledge base and the conversation specification would be interchangeable with the specification and knowledge base of another medication. This would mean that the system could provide counseling for several types of medication, depending on the specification and knowledge base received.

\subsection{Conversationalist}

The Conversationalist is the agent that converses directly with the customer and is therefore the customer-facing part of the system. The agent follows the strict procedure outlined by the YAML document, while answering questions and asking follow-up questions if the answers from the customer are unclear. Follow-up (context-refining) questions, also known as iterative retrieval \cite{huang_survey_2025}, help LLMs refine their situational understanding. However, given the propensity of LLMs to take user input at face value, it is often difficult to make them ask these questions. In our case, the LLM on its own lacks the contextual information it needs to determine the relevance of a given allergy or contraindication. A way to provide more context is by giving the LLM access to a tool that provides responses it can act upon, following the structure: ``if the tool returns X, ask the customer Y''. The aim is that by doing this, the LLM can better determine if a follow-up question should be asked, an approach that seems promising. Finally, few-shot prompts, or examples of how to perform a task, can be very helpful to an LLM \cite{brown_language_2020}. For us, this seems like a fruitful strategy in instructing the Conversationalist how it should confirm customer's answers and re-confirm if the customer changes their answer. Other use cases for it include inter-agent communication and tool usage examples. Providing an agent with examples can also help enforce the habit of asking follow-up questions.

Some of the steps are more critical than others which the YAML document also specifies by a stating a risk level. One of the most critical steps is checking for possible contraindications, which includes allergies, underlying diseases or regular medications. The Conversationalist's final task is to tell the customer which ECPs, if any, can be consumed. It is vital that this agent follows the conversation specification and that it relays information within the system without losing parts of or altering the content. We aim to lessen this risk by having the agent know little about the other agents in the system; it simply knows what to provide as input to each agent, the agent's purpose and what to expect as output, which retains more space in the context window for core functionality (e.g., tasks and responsibilities) and avoids polluting the context with superfluous information.

The agent was also given a name, making it easier to write less ambiguous instructions. Instead of instructions in the more ambiguous form of "You should" it could now be instructed by name, "[name] should".

\subsection{Medicine Interpreter}

{\sloppy

The Medicine interpreter has the task of interpreting the list of the customer's contraindications and, based on those, retrieving a list of ECPs that are safe to consume. The agent extracts individual terms from mentioned contraindications and compares them with the knowledge base using the tools \emph{check\_pill\_contraindicating\_allergies} and \emph{check\_pill\_contraindicating\_medications\_and\_diseases}. Before these tools, pure inference was used to extract the terms which was used as arguments, but this often led to incorrect behavior. For example, the system was insufficient in handling several similar matches to the input term (e.g., potato or corn starch). In cases like these, the agent would choose the first matched term even if there were several, no term at all, or simply re-use the original input term as an argument in the tool call. In an effort to remedy this, the tools \emph{find\_most\_similar\_word\_regular\_medications\_and\_diseases} and \emph{find\_most\_similar\_word\_allergies} were introduced. The tools take an input term and return the most similar word or words from their respective tables. If several terms are returned, the agent should ask the customer a follow-up question for further specification. Furthermore, the agent should ask follow-up questions if a medication contains an ingredient that would affect someone with a hypersensitivity. For example, if the customer says that they are lactose intolerant and some ECPs contain lactose monohydrate, the agent should ask if the customer is severely lactose intolerant, and if so consider those ECPs unsuitable for the customer.

}

The contraindications table of the knowledge base includes disease categories, such as ``Severe Malabsorption Disorder (e.g., Crohn's)'', and conditional contraindications such as ``Asthma (if glucocorticoids)''. It is unlikely that a pharmacy customer suffering from asthma would describe it as ``Asthma, for which I take medication containing glucocorticoids'', or describe their Celiac or Crohn's disease as ``Severe Malabsorption Disorder''. This is a key aspect of the system that demands the flexibility and adaptability of LLMs, as pre-defined states are not able to capture this level of complexity. This means that identifying contraindications in need of re-labeling and asking follow-up questions about conditional contraindications also fell on the Medicine interpreter. As a consequence, it seemed that the agent became overburdened which resulted in a drop in response quality. Seeing as the process consists of two major parts: classification and clarification of conditions and identification of viable ECPs, it is natural to split these tasks between two agents. In summary, the Symptom assessor agent assesses a customer's symptoms and creates a list of contraindications; then the Medicine interpreter uses this list to find ECPs that are safe for the customer to consume.

\subsection{Symptom Assessor Agent} 
\label{symptom_assessor_agent}

The Symptom assessor identifies, matches and re-labels terms, poses follow-up questions if necessary (including contraindications customers might not think of such as breastfeeding) and compiles a list of relevant contraindications. The Medicine interpreter agent expects the input list of contraindications to have irrelevant terms removed and relevant terms matched to the knowledge base. 

The tools \emph{find\_most\_similar\_word\_regular\_medications\\\_and\_diseases} and \emph{find\_most\_similar\_word\_allergies} were given to the Symptom assessor, but were not removed from the Medicine interpreter, as a kind of safeguard, in the event that the Symptom assessor makes an error. The tools have a threshold parameter that dictates how similar a term has to be to constitute a match. The Symptom assessor calls them with the default value of $0.6$, whereas the Medicine interpreter calls them with a threshold of $0.8$ to ensure that the terms are correct event if a small error occurs, such as ``astma'' being input instead of ``asthma.'' While it would be possible to rely on pure inference to get the correct terms, that presents a point of failure, so it should be avoided.

The Symptom assessor seemed to struggle with being tasked to do everything from determining if a disease needed to be re-labeled to forming the final list of relevant terms. Therefore, it was logical to move the potential term re-labeling to its own tool, \emph{classify\_contraindication}: a very simple call to an agent in the form of ``if the input term X is a type of Y, then output Y else None.'' This means, for example, that the input ``Celiac disease'' would produce the output ``Severe Malabsorption Disorder.''

\section{Discussion}

Rule-based systems are often too rigid for contexts where understanding nuance is crucial, such as medication counseling. Pure inference, on the other hand, is too unreliable, meaning that a balance between inferred and rule-based output must be achieved. We present a system designed to achieve such a balance. By subdividing the main task between a network of agents equipped with tools, it aims to increases the reliability of the system, something of utmost importance in a context where errors may have serious consequences. Furthermore, the design of the system is modular in nature, meaning that the system eases the adaption to new situations compared to a single-agent solution.

However, some apparent difficulties with designing this kind of system are still present. Hallucinations are particularly problematic because LLMs rarely admit to lacking knowledge or being wrong; instead they often produce confidently incorrect responses \cite{bubeck_sparks_2023}. This kind of behavior was, for example, prominent when discrepancies between the specification and actual conversation would arise, potentially resulting in some step being skipped.

It is possible some of these difficulties could be mitigated by introducing additional agents, since there currently exists some overlap in the tasks delegated between them. Doing this, however, would require further examination of how to best organize multiple agents in this kind of system. Further examination is here needed.

But whatever strategy is implemented to mitigate potential errors, the need of testing remains when developing a system of this kind; deployment without further supervision involves considerable risk, given the complexities involved in human-agent interaction. Therefore, it would be advisable to implement some kind of adaptive monitoring, starting with a human-in-the-loop upon initial implementation and gradually scaling back the monitoring as the system proves its capabilities.

\subsection{Evaluation}

Even if evaluating a system of this kind is sorely needed, it is difficult. There are metrics, such as whether the customer received the correct recommendation, but it also matters how the system came to that conclusion and how the system interacted with the customer during the process. Based on observations made during the development of the system, some key scenarios were identified and act as situations that the system should be able to handle. These scenarios present varying degrees of ambiguity and should all be manageable by the system. It is crucial that the system can handle more far-fetched descriptions of conditions, not just the normal ones. However, this alone is not a sufficient form of evaluation.

Traditional natural language generation evaluation metrics such as BERTscore \cite{zhang_bertscore_2020} or BARTscore \cite{yuan_bartscore_2021} fall short for different reasons. BERTscore favors token-level similarity and does not consider fluency or coherence \cite{malin_review_2024}. Additionally, it does not consider factuality, which is problematic in a system where that is the chief concern. In contrast, BARTscore suffers from insensitivities and biases when evaluating challenging texts \cite{gao_llm-based_2025}. Furthermore, static evaluations such as these are concerned primarily with evaluating single-turn inputs and outputs \cite{ibrahim_beyond_2024}, as opposed to multi-turn conversations. Due to these shortcomings, another method of evaluation must be used.

A possible alternative is human interaction evaluations (HIE) \cite{ibrahim_beyond_2024}, a category of evaluations that focuses on evaluating the process and outcomes of humans interacting with AI systems. The HIE design framework, as proposed by Ibrahim et al. \cite{ibrahim_beyond_2024} consists of three stages: identifying risk or harm area, characterizing the use context and choosing the evaluation parameters. Naturally, this approach is well-suited for examining the system of this work, as it better captures the complexities of human-AI interactions that more static evaluation methods might overlook.

\subsection{Ethical and Legal Considerations}

Another thing to consider are the ethical considerations within the context of this work --- a central one being the transparency of the system \cite{chow2025large}. The customer has the right to know that they are interacting with an AI-based system \cite{euAIRegulation}, but it is not obvious how transparent it must be \cite{a_bhaskara_agent_2020}. A too transparent system quickly becomes too technical for a lay person, whereas an opaque system builds mistrust and anxiety \cite{quinn_three_2022, xu_medical_2024}. Here, too, balance is key. 

For the sake of transparency, our system could could offer a summarization transcript of the conversation, regarding how the system determined which ECPs to recommend. The transcript could also be used as a means to keep a human-in-the-loop by having domain expertise validate the decision, which could be done in a natural way at the checkout desk. But, as noted, it is not clear exactly what information should be listed here.

Given the sensitive context in which this system is intended to operate, accountability, explainability and liability must also be prioritized \cite{chow2025large}. In the event that an error occurs, there must be a responsible party, as it is not acceptable that the customer is left without recourse for errors. The more autonomous AI systems become, the less clear the question of accountability becomes. The issue is far from clear and is actively being debated \cite{zech_liability_2021, saenz_autonomous_2023, padovan_black_2023, gabison_inherent_2025}.

\section{Conclusion}

We present a proof of concept that leverages LLM agents to provide medication counseling services to customers regarding ECPs, examining narrower and longer interactions than earlier research using LLMs in a pharmacy context. Our design aims through proper constraint of the problem space, task decomposition, use of multiple agents, and tools to increase the level of determinism while maintaining dynamic conversational abilities in the system. We also propose a method of testing and evaluating the system, based on HIE. There still remains much work to be done in designing a system of this kind. Future work should especially focus on finding design patterns that organize multiple agents in ways that make them more reliable. 

Systems that leverage LLMs are likely to become more common going forward, which means that it is already vital to design with an emphasis on reliability and quality. We have shown our approach of doing this in the context of medication counseling with insights relevant to any LLM-based system that needs to balance between flexibility and rigidity.

\bibliographystyle{unsrt}
\bibliography{refs}

\section{Online Resources}
https://github.com/JoarSabel/medical-counselling-llm

\section{Funding}
This work was funded by the Finnish Work Environment Fund and the Swedish Cultural Foundation in Finland.

\end{document}